\documentstyle[12pt]{article}
\begin{document}
 \title{4-Spinors and a Projection onto 3+1 Spacetime}
 \author{Francesco Antonuccio\footnote{The author is trained as a theoretical physicist and is currently a portfolio manager in a London hedge fund.} \\
 London, United Kingdom\\
\texttt{f\_antonuccio@yahoo.co.uk}}
 \renewcommand{\today}{June 12th, 2012}
 \maketitle
 \abstract 
 We write down an explicit projection that maps any given 4-spinor to a point in 3+1 spacetime while commuting with the Lorentz action. This suggests that a Lorentz invariant theory - including spacetime itself - has a more natural expression in terms of these primitive spinor variables, while an ordinary spacetime interpretation may be obtained by projecting solutions. Using this projection, we show how the real components of a given 4-spinor reference a point in a five dimensional spacetime.  \\
\section {Introduction}
In this article we write down an explicit projection that maps a given 4-spinor to a spacetime point in 3+1 dimensions while commuting with the Lorentz action. More precisely, if $L$ is a Lorentz transformation acting on 4-spinors, and $L^{'}$ is the same transformation but in a 3+1 spacetime representation (i.e. acting on spacetime points), then there exists a projection map $P$ from the space of 4-spinors to 3+1 spacetime satisfying
\begin{equation}
P \cdot (L \cdot \Psi) = L^{'} \cdot (P \cdot \Psi)   
\end{equation}
for all 4-spinors $\Psi$. It is in this sense that we may identify a 4-spinor with a corresponding point in spacetime. However, since the mapping is a projection, there are degrees of freedom contained in the spinor space that are hidden from ordinary spacetime. 

The existence of such a projection opens up the possibility of formulating dynamical theories on spacetime - or of spacetime itself - that have their most natural expression in terms of spinor variables. We remark that the twistor theory program is also motivated by an attempt to treat spacetime as derived from more fundamental spinor degrees of freedom \cite{Roger}, but the approach adopted here is not reliant on the special properties of complex numbers, and can be applied to any number of spacetime dimensions. 

Interestingly, the projection map that we analyse has a natural embedding in five dimensional spacetime, and so the eight real components of a given 4-spinor encode information about a corresponding point in five dimensional space. 
The idea that a hidden fifth dimension could be useful in physics has appeared in a number of different contexts. The fifth dimension was perhaps first popularised in Kaluza-Klein models, but in more recent times, ideas involving warped spacetimes \cite{RS}, extra dimensions \cite{Nima}, and even space-time-matter \cite{STM} have emerged from the physics community.  While the importance of the fifth dimension is a serious topic of investigation, we restrict our attention to proving the existence of projection maps from spinor space to spacetime that preserve Lorentz invariance. A nice way to achieve this is to study representations of the Lorentz symmetry over the hyperbolic numbers.   

Although hyperbolic numbers and spinor representations have been discussed before \cite{Hucks},\cite{Ant1},\cite{Ulrych}, the results here are new. For completeness, we will attempt to provide a self-contained account of all the relevant mathematical details at the expense of repeating some basic properties of hyperbolic numbers and representations of the Lorentz group over this number system. 
 
With this in mind, we begin in Section 2 with an introduction of the hyperbolic number system \cite{HyperbolicHist}, and then follow by defining hyperbolic unitary groups, which are the natural analog of complex unitary groups.  

We then proceed in Section 3  by constructing an explicit representation of the Lorentz group for 3+1 dimensional spacetime in terms of $4 \times 4$ matrices defined over the hyperbolic number system. 

In Section 4 we introduce two distinct projection maps that relate any given (hyperbolic) 4-spinor to a corresponding point $(x^0,x^1,x^2,x^3)$ in 3+1 dimensional spacetime. These projection maps have the special property that they commute with the Lorentz action.  Closer investigation reveals that these projection maps are naturally embedded in five dimensional spacetime. 

It is important to remark that hyperbolic 4-spinors are equivalent to Dirac 4-spinors after making a suitable identification of the eight real components \cite{Ant1}. Consequently, any explicit map from a hyperbolic 4-spinor to a spacetime point yields an equivalent map defined over the space of Dirac 4-spinors.

Finally, in Section 5 we give a summary of our results, and speculate on possible applications.

\section {Hyperbolic Numbers and Special Unitary Group SU$(n,{\bf D})$}
The basic properties of the hyperbolic numbers is well documented \cite{HyperbolicHist}. Only the definitions and results contained in the next two subsections are required to understand the results presented in this paper.

\subsection{The Hyperbolic Numbers}
We consider numbers of the form 
\begin{equation}
w = x+{\rm j}y
\end{equation}
where $x$ and $y$ are real numbers, and ${\rm j}$ is a commuting element satisfying the identity
\begin{equation}
{\rm j}^2=1.
\end{equation}
We will denote the set of all such numbers by the symbol ${\bf D}$. Addition, subtraction, and multiplication are defined in the obvious way:
\begin{equation} 
\begin{array}{rcl}
(x_{1}+{\rm j}y_{1}) \pm (x_{2}+{\rm j}y_{2})   & = & ( x_{1}\pm x_{2}) + {\rm j}  ( y_{1}\pm y_{2}) , \\
(x_{1}+{\rm j}y_{1}) \cdot (x_{2}+{\rm j}y_{2})  & = &  (x_{1}x_{2}+y_{1}y_{2})+{\rm j} (x_{1}y_{2}+y_{1}x_{2}). 
\end{array}
\end{equation} 
Moreover, given any hyperbolic number $w = x+{\rm j}y$, we define the `conjugate of $w$', written $\overline{w}$, to be
\begin{equation}
\label{eq: wbar}
\overline{w} \equiv x-{\rm j}y.
\end{equation}
It is now easy to check that the following properties are true for all $w_{1},w_{2} \in {\bf D}$:
\begin{equation}
\begin{array}{c}
\overline{w_{1}+w_{2}}  =  \overline{w_{1}}+\overline{w_{2}}, \\
\overline{w_{1} \cdot w_{2}}  =  \overline{w_{1}} \cdot \overline{w_{2}}.
\end{array}
\end{equation}
It also follows that 
\begin{equation}
\overline{w} \cdot w = x^2-y^2
\end{equation}
for any hyperbolic number $w = x+{\rm j}y$. Thus $\overline{w} \cdot w$ is always real valued, although unlike the case for complex numbers, it may take on negative values. Despite this, we will find it convenient to define the `modulus squared' of $w$, written ${| w |}^{2}$, as
\begin{equation}
{| w |}^2 \equiv \overline{w} \cdot w .
\end{equation}
A nice consequence of these definitions is that for any hyperbolic numbers  $w_{1},w_{2} \in {\bf D}$, we have 
\begin{equation}
{| w_{1} \cdot w_{2} |}^2 = {| w_{1} |}^2 \cdot {| w_{2} |}^2.
\end{equation}
Note that if ${| w |}^2$ is nonzero, then the quantity $w^{-1}$ defined by
\begin{equation}
w^{-1} \equiv \frac{1}{{| w |}^{2}} \cdot \overline{w}
\end{equation}
is a well-defined and unique inverse for $w$. So $w \in {\bf D}$  fails to have an inverse if and only if ${| w |}^2 = x^2-y^2 = 0$. The hyperbolic number system is therefore a non-division algebra.

Note that if $\theta$ is any real parameter, then
\begin{equation}
e^{{\rm j} \theta} = {\rm cosh}\theta + {\rm j} {\rm sinh} \theta .
\end{equation}
If follows that 
\begin{equation}
| e^{{\rm j} \theta}|^2 = 1.
\end{equation}
Moreover, if $w = x + {\rm j} y$ is any given hyperbolic number, then 
\begin{equation}
|e^{{\rm j} \theta} w |^2 = |w|^2,
\end{equation}
and so the quantity $e^{{\rm j}\theta}$ may be viewed as a hyperbolic `phase factor' which preserves the modulus of a hyperbolic number under multiplication. 

Finally, in analogy with the complex numbers, we define the `real' and `imaginary' parts of a given hyperbolic number $w = x + {\rm j}y$, written ${\it Re} \left[ w \right]$ and ${\it Im} \left[ w \right]$ respectively, in the obvious way:
\begin{equation}
\begin{array}{cc}
{\it Re} \left[ w \right] = x, &  {\it Im} \left[ w \right] = y.
\end{array}
\end{equation}

\subsection{The Hyperbolic Unitary Groups} 
Suppose $H$ is an $n \times n$ matrix defined over the hyperbolic numbers ${\bf D}$.  Then the conjugate transpose of $H$, written $H^{\dag}$,  is defined by conjugating each of the elements in $H$ in accordance with definition (\ref{eq: wbar}), and then transposing the resulting matrix:
\begin{equation}
 H^{\dag} \equiv {\overline{H}}^{T}. 
\end{equation}
We say $H$ is Hermitian with respect to ${\bf D}$ if 
$H^{\dag} = H$, and anti-Hermitian if $H^{\dag} = -H$.

Note that if $H$ is an $n \times n$ Hermitian matrix over ${\bf D}$, then $U \equiv e^{{\rm j}H}$ has the property 
\begin{equation}
\label{eq: unitary}
U^{\dag} \cdot U = U \cdot U^{\dag} = 1.
\end{equation}
The set of all $n \times n$ matrices over ${\bf D}$ satisfying the above constraint forms a group manifold, which we will denote as U$(n,{\bf D})$, and call the `unitary group of $n \times n$ matrices over ${\bf D}$', or simply `hyperbolic unitary group'. The special hyperbolic unitary group denoted by SU$(n, {\bf D})$ is a subgroup of U$(n,{\bf D})$, and is defined as all elements $U \in$ U$(n,{\bf D})$ that satisfy the additional constraint
\begin{equation}
{\rm det} U = 1.
\end{equation}
Note that if $\Psi$ denotes an $n$-component column vector over ${\bf D}$ (i.e. $\Psi \in {\bf D}^n)$, then we may write
\begin{equation}
\Psi \equiv \left(
\begin{array}{c}
 w_{1} \\
 w_{2} \\
\vdots \\
w_{n}
\end{array}
\right),
\end{equation} 
which implies
\begin{equation}
\label{eq: psimodsquared}
\Psi^{\dag} \Psi = |w_{1}|^2+|w_{2}|^2+ \cdots + |w_{n}|^2.
\end{equation}
In other words, $\Psi^{\dag} \Psi$ is just a real number, and can take on negative values. For notational convenience, we will sometimes write $|\Psi|^2$ instead of  $\Psi^{\dag} \Psi$:
\begin{equation}
\label{eq: modulus}
|\Psi|^2 \equiv \Psi^{\dag} \Psi.
\end{equation}
 It now follows from definitions (\ref{eq: unitary})  and (\ref{eq: modulus}) above that  
\begin{equation}
|U \cdot \Psi|^2 = |\Psi|^2 
\end{equation}
for any $U \in$ SU$(n, {\bf D})$ and any $n$-component column vector $\Psi \in {\bf D}^n$. In other words, for any $\Psi \in {\bf D}^n$, the simple scalar quantity  $|\Psi|^2$ given by equation (\ref{eq: psimodsquared}) is invariant under SU$(n, {\bf D})$ transformations.

In the next Section, we show that the Lorentz group in 3+1 dimensions is a subgroup of SU$(4, {\bf D})$, where group elements act on a 4-component spinor $\Psi$. From the discussion above, it follows that the quantity $|\Psi|^2$ is a Lorentz invariant scalar. 

\section{Representations of the Lorentz Group}
We now present explicit representations of the Lorentz group on 3+1 spacetime, along with the associated Lie algebra. 

\subsection{The Standard Lorentz Group on 3+1 Spacetime}
Given coordinates $(x^0,x^1,x^2,x^3)$ in 3+1 spacetime with metric signature $(+---)$, a Lorentz transformation on these coordinates corresponds to any  linear map (i.e. $4 \times 4$ matrix $L$)  acting on the four coordinates of spacetime,
\begin{equation}
\label{eq: lorentzmap}
\begin{array}{ccc}
\left(
\begin{array}{c}
x^0 \\
x^1 \\
x^2 \\
x^3
\end{array}
\right)

& \rightarrow &

L \cdot \left(
\begin{array}{c}
x^0 \\
x^1 \\
x^2 \\
x^3
\end{array}
\right),

\end{array}
\end{equation}
which preserves the quantity
\begin{equation}
\label{eq: 4Dlorentzscalar}
(x^0)^2-(x^1)^2-(x^2)^2-(x^3)^2.
\end{equation} 
These conditions imply that if $L$ is continuoulsy connected to the identity, it must take the form
\begin{equation}
\label{eq: realLorentztransform}
L = e^{\alpha_1 E_1+\alpha_2 E_2+\alpha_3 E_3+\beta_1 F_1+\beta_2 F_2 +\beta_3 F_3 }
\end{equation} 
where the $\alpha_i$ and $\beta_i$ are real numbers and $E_i$,$F_i$ are real matrices defined as follows:
\begin{equation}
\begin{array}{ccc}
E_1 = \left(
\begin{array}{cccc}
0 & 0 & 1 & 0 \\
0 & 0 & 0 & 0 \\
1 & 0 & 0 & 0 \\
0 & 0 & 0 & 0
\end{array}
\right)
&
E_2 = \left(
\begin{array}{cccc}
0 & 0 & 0 & 0 \\
0 & 0 & 0 & 0 \\
0 & 0 & 0 & 1 \\
0 & 0 & -1 & 0
\end{array}
\right)
&
E_3 = \left(
\begin{array}{cccc}
0 & 0 & 0 & 1 \\
0 & 0 & 0 & 0 \\
0 & 0 & 0 & 0 \\
1 & 0 & 0 & 0
\end{array}
\right)
\\
\\
F_1 = \left(
\begin{array}{cccc}
0 & 0 & 0 & 0 \\
0 & 0 & 0 & -1 \\
0 & 0 & 0 & 0 \\
0 & 1 & 0 & 0
\end{array}
\right)
 & 
F_2 = \left(
\begin{array}{cccc}
0 & -1 & 0 & 0 \\
-1 & 0 & 0 & 0 \\
0 & 0 & 0 & 0 \\
0 & 0 & 0 & 0
\end{array}
\right)
& 
F_3 = \left(
\begin{array}{cccc}
0 & 0 & 0 & 0 \\
0 & 0 & 1 & 0 \\
0 & -1 & 0 & 0 \\
0 & 0 & 0 & 0
\end{array}
\right)

\end{array}
\end{equation}
From a physical point of view, $E_1$, $F_2$ and $E_3$ correspond to Lorentz boosts parallel to the three spatial directions, while $F_1$, $E_2$ and $F_3$ correspond to rotations about the three spatial coordinate axes. It is in this sense that there are six generators for the Lorentz group on 3+1 spacetime.  

By direct substitution, one can check that these six generators of the Lorentz group satisfy the following commutation relations:
\begin{equation}
\begin{array}{llll}
\label{eq: 4Dliealgebra}
[E_1,E_2] = E_3  &  [F_1,F_2] = -E_3  &  [E_1,F_2] = F_3  &  [F_1,E_2] = F_3  \\
\left[E_2,E_3 \right] = E_1  &  \left[F_2,F_3 \right] = -E_1  &  \left[ E_2,F_3 \right] = F_1  &  \left[ F_2,E_3 \right] = F_1 \\
\left[ E_3,E_1 \right] = -E_2  &  \left[ F_3,F_1 \right] = E_2  &  \left[ E_3,F_1 \right] = -F_2  &  \left[ F_3,E_1 \right] = -F_2 
\end{array}
\end{equation}

All other commutators vanish. Abstractly, these relations define the Lie Algebra of the Lorentz group on 3+1 spacetime.

If we now view the real coeeficients $\alpha_i, \beta_i$ appearing in (\ref{eq: realLorentztransform}) as infinitesimally small, then the transformation (\ref{eq: lorentzmap}) on the spacetime coordinates $(x^0,x^1,x^2,x^3)$ takes the following explicit form:
\begin{equation}
\label{eq: x0map}
x^0 \rightarrow x^0 -\beta_2 x^1+\alpha_1 x^2 +\alpha_3 x^3
\end{equation}
\begin{equation}
x^1 \rightarrow x^1 -\beta_2 x^0+\beta_3 x^2 -\beta_1 x^3
\end{equation}
\begin{equation}
x^2 \rightarrow x^2 +\alpha_1 x^0-\beta_3 x^1 -\alpha_2 x^3
\end{equation}
\begin{equation}
\label{eq: x3map}
x^3 \rightarrow x^3 +\alpha_3 x^0+\beta_1 x^1 -\alpha_2 x^2
\end{equation}
where we have dropped all terms beyond the linear approximation.

In the next Section, we find another representation of the Lorentz Lie algebra (\ref{eq: 4Dliealgebra}) in terms of $4 \times 4$ anti-Hermitian matrices defined over the hyperbolic numbers ${\bf D}$. 

\subsection{Hyperbolic Representation of the 3+1 Lorentz Group}
\label{lorentzinvariance}
Our objective in this section is to find $4 \times 4$ anti-Hermitian matrices over ${\bf D}$ that satisfy the Lorentz Lie algebra (\ref{eq: 4Dliealgebra}). We begin by defining three $2 \times 2$ anti-Hermitian matrices $\tau_1$, $\tau_2$ and $\tau_3$ as follows:
\begin{equation}
\begin{array}{ccc}
\tau_1 = \frac{1}{2}
\left(
\begin{array}{cc}
0 & {\rm j} \\
{\rm j} & 0
\end{array}
\right)
&
\tau_2 = \frac{1}{2}
\left(
\begin{array}{cc}
0 & -1 \\
1 & 0
\end{array}
\right)
&
\tau_3 = \frac{1}{2}
\left(
\begin{array}{cc}
{\rm j} & 0 \\
0 & -{\rm j}
\end{array}
\right)
\end{array}
\end{equation}
It follows that these matrices satisfy the following commutation relations:
\begin{equation}
\begin{array}{ccc}
\left[ \tau_1, \tau_2 \right] = \tau_3  &  \left[ \tau_2, \tau_3 \right] = \tau_1 & \left[ \tau_3, \tau_1 \right] = -\tau_2 .
\end{array}
\end{equation}
We are now ready to define the $4 \times 4$ matrices $E_i$ and $F_i$ that will form the basis of our hyperbolic representation:
\begin{equation}
\label{eq: hyperbolicrep}
\begin{array}{ccc}
E_1 = \left(
\begin{array}{cc}
\tau_1 & 0 \\
0 & \tau_1
\end{array}
\right)
&
E_2 = \left(
\begin{array}{cc}
\tau_2 & 0 \\
0 & \tau_2
\end{array}
\right)
&
E_3 = \left(
\begin{array}{cc}
\tau_3 & 0 \\
0 & \tau_3
\end{array}
\right)
\\ \\
F_1 = {\rm j} \left(
\begin{array}{cc}
 0 & \tau_1 \\
-\tau_1 & 0
\end{array}
\right)
&
F_2 = {\rm j} \left(
\begin{array}{cc}
 0 & \tau_2 \\
-\tau_2 & 0
\end{array}
\right)
&
F_3 = {\rm j} \left(
\begin{array}{cc}
 0 & \tau_3 \\
-\tau_3 & 0
\end{array}
\right)
\end{array}
\end{equation}
It is straightforward to check that the matrices  $E_i$ and $F_i$ defined above are indeed anti-Hermitian with respect to ${\bf D}$  (i.e. $E_{i}^{\dag} = -E_i$ and $F_{i}^{\dag} = -F_i$), and satisfy the 3+1 Lorentz Lie algebra of commutation relations (\ref{eq: 4Dliealgebra}). 
Consequently, a Lorentz transformation $L$ in the representation specified by the generating matrices (\ref{eq: hyperbolicrep}) takes the form 
\begin{equation}
\label{eq: realLorentztransformII}
L = e^{\alpha_1 E_1+\alpha_2 E_2+\alpha_3 E_3+\beta_1 F_1+\beta_2 F_2 +\beta_3 F_3 }
\end{equation} 
where the $\alpha_i$ and $\beta_i$ are real numbers as before, but $L$ now acts on 4-component hyperbolic spinors $\Psi \in {\bf D}^4$:
\begin{equation}
\label{eq: explicithyperbolictransform}
 \left(
 \begin{array}{c}
a_1+ \rm{j} b_1 \\
a_2+ \rm{j} b_2 \\
a_3+ \rm{j} b_3 \\
a_4+ \rm{j} b_4
\end{array}
\right) \rightarrow
L \cdot 
 \left(
 \begin{array}{c}
a_1+ \rm{j} b_1 \\
a_2+ \rm{j} b_2 \\
a_3+ \rm{j} b_3 \\
a_4+ \rm{j} b_4
\end{array}
\right) .
\end{equation}
The components $a_i$ and $b_i$ appearing above are real numbers. Since the $E_i$ and $F_i$ appearing in  (\ref{eq: realLorentztransformII}) are anti-Hermitian, then 
\begin{equation}
L^\dag = e^{-\left(\alpha_1 E_1+\alpha_2 E_2+\alpha_3 E_3+\beta_1 F_1+\beta_2 F_2 +\beta_3 F_3\right) } =L^{-1}.
\end{equation}
In other words $L^{\dag}$ is the inverse of $L$:
\begin{equation}
\label{eq: inversecondition}
L^{\dag} L = L L^{\dag} = 1.
\end{equation}
Moreover, the generating matrices $E_i$ and $F_i$  are traceless, so we have the additional property
\begin{equation}
\label{eq: detcondition}
{\rm det} L = 1.
\end{equation}
It follows from (\ref{eq: inversecondition}) and  (\ref{eq: detcondition}) above that the Lorentz transformation $L$ defined by  (\ref{eq: realLorentztransformII}) is an element of the special hyperbolic unitary group SU$(4,{\bf D})$. This group manifold actually has fifteen generators in its Lie algebra, and so we conclude that the Lorentz group on 3+1 spacetime is a subgroup of SU$(4,{\bf D})$. 

If we assume the real coefficients $\alpha_i$ and $\beta_i$ appearing in (\ref{eq: realLorentztransformII})  are infinitesimally small, then up to linear order in these coefficients, we may write the Lorentz transformation  (\ref{eq: explicithyperbolictransform}) explicitly in terms of the eight real components $a_i$ and $b_i$:
\begin{equation}
\label{eq: a1map}
a_1 \rightarrow a_1 +\frac{1}{2} \left(
\alpha_1 b_2 -\alpha_2 a_2 + \alpha_3 b_1 + \beta_1 a_4 - \beta_2 b_4 + \beta_3 a_3
\right)
\end{equation}
\begin{equation}
a_2 \rightarrow a_2 +\frac{1}{2} \left(
\alpha_1 b_1 +\alpha_2 a_1 - \alpha_3 b_2 + \beta_1 a_3 + \beta_2 b_3 - \beta_3 a_4
\right)
\end{equation}
\begin{equation}
a_3 \rightarrow a_3 +\frac{1}{2} \left(
\alpha_1 b_4 -\alpha_2 a_4 + \alpha_3 b_3 - \beta_1 a_2 + \beta_2 b_2 - \beta_3 a_1
\right)
\end{equation}
\begin{equation}
a_4 \rightarrow a_4 +\frac{1}{2} \left(
\alpha_1 b_3 +\alpha_2 a_3 -\alpha_3 b_4 - \beta_1 a_1 - \beta_2 b_1 + \beta_3 a_2
\right)
\end{equation}
\begin{equation}
b_1 \rightarrow b_1 +\frac{1}{2} \left(
\alpha_1 a_2 -\alpha_2 b_2 + \alpha_3 a_1 + \beta_1 b_4 - \beta_2 a_4 + \beta_3 b_3
\right)
\end{equation}
\begin{equation}
b_2 \rightarrow b_2 +\frac{1}{2} \left(
\alpha_1 a_1 +\alpha_2 b_1 - \alpha_3 a_2 + \beta_1 b_3 + \beta_2 a_3 - \beta_3 b_4
\right)
\end{equation}
\begin{equation}
b_3 \rightarrow b_3 +\frac{1}{2} \left(
\alpha_1 a_4 -\alpha_2 b_4 + \alpha_3 a_3 - \beta_1 b_2 + \beta_2 a_2 - \beta_3 b_1
\right)
\end{equation}
\begin{equation}
\label{eq: b4map}
b_4 \rightarrow b_4 +\frac{1}{2} \left(
\alpha_1 a_3 +\alpha_2 b_3 - \alpha_3 a_4 - \beta_1 b_1 - \beta_2 a_1 + \beta_3 b_2
\right)
\end{equation}

The infinitesimal Lorentz transformations on real spacetime points given by (\ref{eq: x0map})-(\ref{eq: x3map}), and on hyperbolic spinors specified by (\ref{eq: a1map})-(\ref{eq: b4map})  above will be important in the next section. In particular, we show there exists a projection that maps the spinor components $a_i$ and $b_i$ to a real spacetime point $(x^0,x^1,x^2,x^3)$  that respects simultaneously the relations  (\ref{eq: x0map})-(\ref{eq: x3map}) and (\ref{eq: a1map})-(\ref{eq: b4map}) under an infinitesimal Lorentz transformation. More succinctly, there exists a projection map that commutes with the Lorentz transformation.

\section{A Projection from Spinors to Spacetime}
In the following, we provide two examples of projection maps from hyperbolic spinors to 3+1 spacetime that commute with the Lorentz transformation. The first projection has a natural embedding in 4+1 spacetime, while the second has a natural embedding in 3+2 spacetime (i.e. three spatial and two time directions). Despite these striking differences, the two mappings are in fact closely related to each other.
\subsection{A Projection with 4+1 Spacetime Embedding}
Suppose we are given a hyperbolic spinor $\Psi \in {\bf D}^4$. Then we may write 
\begin{equation}
\label{eq: spinor}
\Psi = \left(
 \begin{array}{c}
a_1+ \rm{j} b_1 \\
a_2+ \rm{j} b_2 \\
a_3+ \rm{j} b_3 \\
a_4+ \rm{j} b_4
\end{array}
\right),
\end{equation}
where the eight components $a_i$ and $b_i$ are real numbers. We now assign to each hyperbolic spinor (\ref{eq: spinor}) a point $(x^0,x^1,x^2,x^3)$  in 3+1 dimensional spacetime by making the following identification:
\begin{equation}
\label{eq: projection}
\begin{array}{l}
x^0 = \frac{1}{2}\left(
a_1^2 +a_2^2+a_3^2+a_4^2+b_1^2+b_2^2+b_3^2+b_4^2
\right) \\
x^1 =  a_1 b_4 +a_4 b_1 -a_2 b_3 -a_3 b_2 \\
x^2 =  a_1 b_2 + a_2 b_1 +a_3 b_4 + a_4 b_3 \\
x^3 = a_1 b_1 -a_2 b_2 +a_3 b_3 -a_4 b_4 \\
\end{array}
\end{equation}
It is now straightforward to show that under an infinitesimal Lorentz transformation of the hyperbolic spinor components $a_i$ and $b_i$ specified by (\ref{eq: a1map})-(\ref{eq: b4map}), the spacetime point $(x^0,x^1,x^2,x^3)$  defined by the projection map (\ref{eq: projection}) transforms in exactly the way specified by (\ref{eq: x0map})-(\ref{eq: x3map}).  Repeated applications of these infinitesimal transformations yields - in the limit - the main technical result in this paper. Namely, the projection (\ref{eq: projection}) that maps a given hyperbolic spinor to a point in 3+1 spacetime commutes with the Lorentz action. It is in this sense that we may `identify' points in the spinor space with ordinary points in spacetime via this projection.

To gain further insight, we recall that if $\Psi$ is a hyperbolic spinor, then the quantity $|\Psi |^2 \equiv \Psi^{\dag}\Psi$ is invariant under 3+1 Lorentz transformations (see Section \ref{lorentzinvariance} for details). Using the identifications provided by the projection map  (\ref{eq: projection}), and some simple algebraic manipulation, one arrives at the following identity:

\begin{equation}
\label{eq: 5Dequality}
\left[  \frac{1}{2} \left( \Psi^{\dag} \Psi \right) \right]^2 = (x^0)^2 - (x^1)^2 - (x^2)^2 - (x^3)^2 - (x^4)^2
\end{equation}
where $x^4$ is defined in terms of the spinor components $a_i$ and $b_i$  as follows:
\begin{equation}
\label{eq: x4defn}
x^4 = a_2 b_4 +a_1 b_3 -a_4 b_2 -a_3 b_1.
\end{equation}
One can show that under the infinitesimal Lorentz transformations specified by (\ref{eq: a1map})-(\ref{eq: b4map}), the additional fifth coordinate $x^4$ defined by (\ref{eq: x4defn}) is invariant, and so does not mix with the other spacetime coordinates $x^0,x^1,x^2$, and $x^3$. Nevertheless, the identity (\ref{eq: 5Dequality}) suggests that the projection map (\ref{eq: projection}) is naturally embedded in a 4+1 dimensional spacetime.

If we write $w_i = a_i + {\rm j}b_i$ so that the hyperbolic spinor (\ref{eq: spinor}) takes the equivalent form
\begin{equation}
\label{eq: HypSpinor}
\Psi = \left(
 \begin{array}{c}
w_1 \\
w_2 \\
w_3 \\
w_4
\end{array}
\right),
\end{equation}  
then the projection map defined by (\ref{eq: projection}) and (\ref{eq: x4defn}) takes the following form:
\begin{equation}
\label{eq: HypProjection}
\begin{array}{l}
x^0 = {\it Re} \left[ \frac{1}{2}\left(
w_1^2+w_2^2+w_3^2+w_4^2
\right) \right] \\
x^1 =  {\it Im} \left[ w_1 w_4 - w_2 w_3  \right] \\
x^2 =   {\it Im} \left[ w_1 w_2 + w_3 w_4  \right] \\
x^3 = {\it Im} \left[ \frac{1}{2}\left(
w_1^2-w_2^2+w_3^2-w_4^2
\right) \right] \\
x^4 =   {\it Im} \left[ \overline{w}_2 w_4 + \overline{w}_1 w_3  \right], \\
\end{array}
\end{equation}
where we have included the fifth coordinate $x^4$ for completeness. Note that under the hyperbolic phase transformation $\Psi \rightarrow e^{{\rm j} \theta} \Psi$, this fifth coordinate $x^4$ remains invariant. Since $|\Psi |^2$ is also manifestly invariant under such a transformation, we conclude from identity (\ref{eq: 5Dequality}) that the mapping $\Psi \rightarrow e^{{\rm j} \theta} \Psi$ preserves the metric on 3+1 spacetime under the projection map. We have already shown that the three boosts and three spatial rotations of the Lorentz group generate a subgroup of SU$(4,{\bf D})$ when acting on spinors. The hyperbolic phase transformation discussed above is not an element of this group. Rather, it belongs to the hyperbolic unitary group U$(4,{\bf D})$, since its determinant is in general not equal to one, but nevertheless preserves the quantity $|\Psi |^2$. Obviously, a better understanding of this global boost transformation in terms of the underlying physics is required.

It is worth noting that in the projection defined above, the time coordinate $x^0$ is never negative. It turns out, however, that there exists another projection where the time coordinate is allowed to take on any value, positive or negative. This new projection is nevertheless intimately related to the above projection, and is the subject of the next section.

\subsection{A Projection with 3+2 Spacetime Embedding}
We now define a related projection that maps a given hyperbolic spinor (\ref{eq: spinor}) to a point $(x^0,x^1,x^2,x^3)$ in 3+1 spacetime, and which has a natural embedding in a five dimensional Lorentz spacetime with metric signature $(-+++-)$. The explicit representation of this map is stated below:
\begin{equation}
\label{eq: projectionII}
\begin{array}{l}
x^0 = a_1 b_1 +a_2 b_2 +a_3 b_3 +a_4 b_4 \\
x^1 = a_1 a_4 + b_1 b_4 - a_2 a_3 - b_2 b_3 \\
x^2 = a_1 a_2 + b_1 b_2 + a_3 a_4 + b_3 b_4 \\
x^3 = \frac{1}{2} \left(a_1^2 +b_1^2 -a_2^2-b_2^2 +a_3^2+b_3^2-a_4^2-b_4^2 \right) \\
\end{array}
\end{equation}
 The spinor projection onto 3+1 spacetime defined by (\ref{eq: projectionII}) can be shown to commute with the Lorentz transformation. Equivalently, under infinitesimal transformations of the spinor components $a_i$ and $b_i$ specified by (\ref{eq: a1map})-(\ref{eq: b4map}), the spacetime point $(x^0,x^1,x^2,x^3)$ defined by the relations (\ref{eq: projectionII}) transforms in exactly the way specified by (\ref{eq: x0map})-(\ref{eq: x3map}). 

Using the relations (\ref{eq: projectionII})  between spacetime points and spinor components, one can show that the following identity holds:
\begin{equation}
\label{eq: 5DequalityII}
\left[  \frac{1}{2} \left( \Psi^{\dag} \Psi \right) \right]^2 = -(x^0)^2 + (x^1)^2 + (x^2)^2 + (x^3)^2 - (x^4)^2,
\end{equation}
where the additional fifth coordinate $x^4$ appearing above is defined in terms of the spinor components $a_i$ and $b_i$  as follows:
\begin{equation}
\label{eq: x4defnII}
x^4 = a_2 b_4 +a_1 b_3 -a_4 b_2 -a_3 b_1.
\end{equation}
Note that the above definition for the fifth coordinate $x^4$ exactly coincides with definition (\ref{eq: x4defn}). As before, this implies $x^4$ behaves as an invariant scalar under the infinitesimal Lorentz transformations specified by (\ref{eq: a1map})-(\ref{eq: b4map}), and so does  not couple with the other spacetime coordinates $x^0,x^1,x^2$, and $x^3$. However, unlike the previous projection, the identity (\ref{eq: 5DequalityII}) makes manifest the idea that the projection map (\ref{eq: projectionII}) is naturally embedded in a five dimensional spacetime with metric signature $(-+++-)$.

As before, let us write the hyperbolic spinor $\Psi$ that we are mapping onto spacetime in the form given by (\ref{eq: HypSpinor}) so that $w_i = a_i + {\rm j} b_i$. Then the projection specified by relations (\ref{eq: projectionII}) and (\ref{eq: x4defnII}) takes the following equivalent form:
\begin{equation}
\label{eq: HypProjectionII}
\begin{array}{l}
x^0 = {\it Im} \left[ \frac{1}{2}\left(
w_1^2+w_2^2+w_3^2+w_4^2
\right) \right] \\
x^1 =  {\it Re} \left[ w_1 w_4 - w_2 w_3  \right] \\
x^2 =   {\it Re} \left[ w_1 w_2 + w_3 w_4  \right] \\
x^3 = {\it Re} \left[ \frac{1}{2}\left(
w_1^2-w_2^2+w_3^2-w_4^2
\right) \right] \\
x^4 =   {\it Im} \left[ \overline{w}_2 w_4 + \overline{w}_1 w_3  \right]. \\
\end{array}
\end{equation}
Note the similarity between the above projection and the projection specified by (\ref{eq: HypProjection}). Specifically, for the 3+1 spacetime coordinates $x^0,x^1,x^2$, and $x^3$, the only difference is the interchange between real and imaginary operators ${\it Re} \leftrightarrow {\it Im}$. The fifth coordinate $x^4$, however, is defined identically in both projections. 

Note (as before) that under the hyperbolic phase transformation $\Psi \rightarrow e^{{\rm j} \theta} \Psi$, this fifth coordinate $x^4$ remains invariant, and so induces a Lorentz preserving transformation in 3+1 space-time by virtue of identity  (\ref{eq: 5DequalityII}). 

\section{Conclusions and Final Remarks}
\label{conclusions}
In this article we gave an explicit representation of a projection that maps a given hyperbolic 4-spinor to a point in 3+1 spacetime while commuting with the Lorentz action. Since hyperbolic 4-spinors are equivalent to Dirac 4-spinors after a suitable identification of components \cite{Ant1}, the existence of such a map on the space of hyperbolic spinors automatically gives rise to a projection map for Dirac 4-spinors. 

Such a projection provides scope for formulating dynamical theories on spacetime - or even of spacetime itself - by working exclusively with these spinor variables. We then make contact with the physical world by projecting solutions onto ordinary spacetime. Since the spinor space contains additional degrees of freedom, physical theories formulated in this way might lead to interesting non-trivial dynamics.

We also discovered that the projection has a natural embedding in a five dimensional spacetime. The extra fifth coordinate $x^4$ that emerges from the spinor components transforms as a scalar under Lorentz transformations on 3+1 spacetime, and so does not mix with the usual spacetime coordinates $x^0$, $x^1$, $x^2$ and $x^3$. 

An important observation is that the Lorentz group on 3+1 spacetime is a subgroup of SU$(4,{\bf D})$. In general, this special hyperbolic unitary group preserves the modulus squared $|\Psi |^2$ for a given hyperbolic 4-spinor $\Psi$, and so by virtue of the identities (\ref{eq: 5Dequality}) and (\ref{eq: 5DequalityII}), the group SU$(4,{\bf D})$ must contain as subroups the symmetries SO$(1,4; {\bf R})$ and SO$(3,2; {\bf R})$. Now the group  SU$(4,{\bf D})$ has fifteen generators, which is larger than the ten generators of the two subgoups mentioned above, and so one might guess that SU$(4,{\bf D})$ is just the conformal group SU$(2,2;{\bf C})$. However, it was pointed out in an earlier work  \cite{Ant1} that SU$(4,{\bf D})$ is isomorphic to SO$(3,3; {\bf R})$, which is not isomorphic to the conformal symmetry SO$(2,4; {\bf R})$. 

All of this suggests that the special hyperbolic unitary group SU$(4,{\bf D})$ (or perhaps hyperbolic unitary group U$(4,{\bf D})$ - see next paragraph) may provide a new perspective when studying the fundamental symmetries of spacetime. 

Finally, we observed that the hyperbolic phase transformation of a 4-spinor given by $\Psi \rightarrow e^{{\rm j} \theta} \Psi$ preserves the metric on 3+1 spacetime since it leaves the modulus squared $|\Psi|^2$  and the fifth coordinate $x^4$ invariant. However, since the determinant is not equal to unity (in general), it is not an element of the Lorentz group on 3+1 spacetime. A proper physical explanation of this transformation is lacking at this point.

\end{document}